\documentclass[11pt,twoside]{article}
\usepackage{asp2004}
\usepackage{psfig}
\usepackage{epsf}
\usepackage{graphics}
\usepackage{lscape}
\def\edcomment#1{\iffalse\marginpar{\raggedright\sl#1\/}\else\relax\fi}
\def\gsim{\lower.73ex\hbox{$\sim$}\llap{\raise.4ex\hbox{$>$}}$\,$}
\def\lsim{\lower.73ex\hbox{$\sim$}\llap{\raise.4ex\hbox{$<$}}$\,$}
\def\mpc{$\,h^{-1}\,$Mpc}
\def\%{~per~cent}

\reversemarginpar

\begin{document}
\title{The Large Local Hole in the Galaxy Distribution: The 2MASS Galaxy Angular 
Power Spectrum}
\author{W.J. Frith, P.J. Outram \& T. Shanks}
\affil{Dept. of Physics, Univ. of Durham, South Road, Durham DH1 3LE, UK}

\begin{abstract}
We present new evidence for a large deficiency in the local galaxy distribution situated in the $\approx$4000 deg$^2$
APM survey area. We use models guided by the 2dF Galaxy Redshift Survey (2dFGRS) $n(z)$ as a probe of the underlying large-scale 
structure. We first check the usefulness of this technique by comparing the 2dFGRS $n(z)$ model prediction with the $K$-band and 
$B$-band number counts extracted from the 2MASS and 2dFGRS parent catalogues over the 2dFGRS Northern and Southern declination 
strips, before turning to a comparison with the APM counts. We find that the APM counts in both the $B$ and $K$-bands indicate a 
deficiency in the local galaxy distribution of $\approx$30\% to $z\approx$0.1 over the entire APM survey area. We examine the 
implied significance of such a large local hole, considering several possible forms for the real-space 
correlation function. We find that such a deficiency in the APM survey area indicates an excess of power at large scales over 
what is expected from the correlation function observed in the 2dFGRS correlation
function or predicted from $\Lambda$CDM Hubble Volume mock catalogues. In
order to check further the clustering at large scales in the 2MASS data,
we have calculated the angular power spectrum for 2MASS galaxies. Although
in the linear regime ($l<30$), $\Lambda$CDM models can give a good fit to
the 2MASS angular power spectrum, over a wider range ($l<100$) the power
spectrum from Hubble Volume mock catalogues suggests that scale-dependent
bias may be needed for $\Lambda$CDM to fit. However, the modest increase
in large-scale power observed in the 2MASS angular power spectrum is still not
enough to explain the local hole. If the APM survey area really is 25\%
deficient in galaxies out to $z\approx0.1$, explanations for the disagreement
with observed galaxy clustering statistics include the possibilities that
the galaxy clustering is non-Gaussian on large scales or that the 2MASS
volume is still too small to represent a `fair sample' of the Universe.
Extending the 2dFGRS redshift survey over the whole APM area would resolve
many of the remaining questions about the existence and interpretation of
this local hole.
\end{abstract}
\thispagestyle{plain}

\section{Introduction}
Since the publication of the APM survey \citep{mad} the deficit of galaxies in the 
resulting number counts has remained a troublesome issue \citep[e.g.][]{met}; is this deficiency ($\approx$50\% 
at B=16) due to real features in the galaxy distribution, galaxy evolution or errors in the 
photometry? If the observed deficit were exclusively due to an underdensity in the 
galaxy distribution over $\approx$4000 deg$^2$ in the Southern Galactic Cap (SGC), it would be 
unexpectedly large for our present understanding of large-scale structure. However, 
invoking strong galaxy evolution at low redshifts is also problematic as it requires 
the presence of a high-redshift tail in the $n(z)$ which is not apparent.  

Evidence for the presence of a large zero-point error in the APM photometry has recently
been provided by \citet{bus}, who find an offset of 0.31 magnitudes with respect to 
CCD $B$-band photometry for 725 matched galaxies over 297 deg$^2$ below $b_J$=17.15. Together with 
a small correction to the fainter $B>$17 APM photometry by \citet{met}, the APM counts 
are now more reasonable with only a $\approx$25\% deficit at $B$=16.

Recently, large redshift surveys have also begun to elucidate a possible cause of the low
APM counts. The 2dF Galaxy Redshift Survey \citep[2dFGRS;][]{col}, the Las Campanas Redshift 
Survey \citep[LCRS;][]{she}, the ESO Slice Project \citep[ESP;][]{vet} and the Durham-UKST  
redshift survey \citep[D-UKST;][]{rat} are all situated within the APM survey area and persistently indicate large deficiencies
in the local galaxy distribution. In particular, the 2dFGRS Southern strip indicates 
the presence of a $\approx$30\% deficiency in the local galaxy distribution to $z\approx$0.1.

In order to determine the effect of this observed structure on the number counts, \citet{fri}  
compared optical counts on the 2dFGRS strips along with the corresponding $K$-band counts extracted from the 
2 Micron All Sky Survey \citep[2MASS;][]{jar} second incremental release catalogue, to models 
guided by the 2dFGRS $n(z)$. They found good agreement between the optical and $K$-band counts 
and the expected trends defined by the 2dFGRS $n(z)$, indicating that 
the number counts in the 2dFGRS fields are a direct consquence of real features in the galaxy distribution 
and that low redshift luminosity evolution does not play a significant role. However, due to incompleteness 
in the 2MASS second incremental release, a comparison between the APM survey area number counts and the corresponding 
$K$-band counts could not be made.

Here, we aim to corroborate the agreement found between the number counts and the 2dFGRS $n(z)$ models over 
the 2dFGRS strips and also probe the galaxy distribution over the entire APM survey area using the newly 
completed 2dFGRS and 2MASS data. To investigate the cosmological relevance of the implied under-density, we 
calculate the significance of such a feature in the galaxy distribution assuming certain forms for the 
two-point correlation function. Finally, we calculate the 2MASS galaxy angular power spectrum to determine 
whether the presence of power at large scales might explain the deficiency. This analysis is more fully 
explained in the forthcoming paper \citet{fri1}.

\section{Probing the Local Galaxy Distribution}

\begin{figure}[!ht]
\centerline{\epsfxsize = 3.0in
\epsfbox{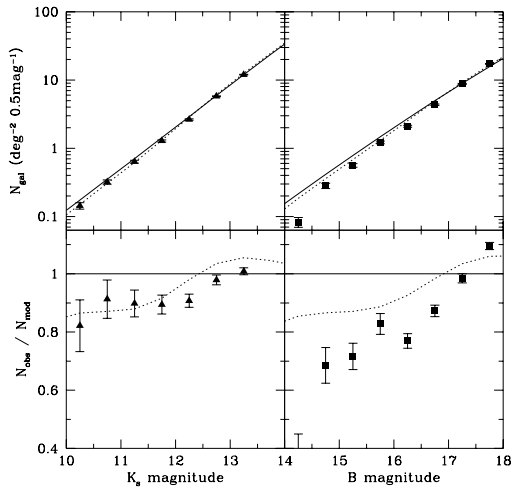}}
\caption{Number magnitude counts in the 2MASS $K_s$-band (left-hand panels) and $B$-band (right-hand panels)
extracted for the 2dFGRS Northern declination strip. The upper panels show the counts on a
logarithmic plot. The lower panels show the residual plot of the counts,
i.e. the number count divided through by the homogeneous prediction. The homogeneous model is indicated by a solid line,
with the model incorporating the 2dFGRS Northern $n(z)$ shown by the dotted line. The errorbars are Poissonian.}
\label{fig:2dfngp}
\end{figure}

\begin{figure}[!ht]
\centerline{\epsfxsize = 3.0in
\epsfbox{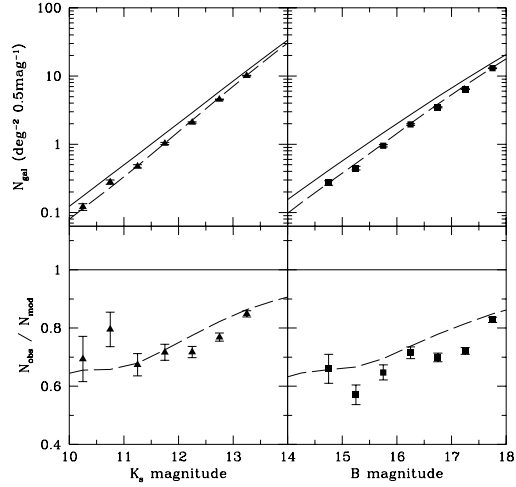}}
\caption{Number magnitude counts in the $K_s$ and $B$-bands for the 2dFGRS Southern declination strip. The panels
are as before, with the homogeneous model again indicated by a solid line, and the model incorporating the 2dFGRS 
Southern $n(z)$ shown by a dashed line. Again, the errors are Poissonian.}
\label{fig:2dfsgp}
\end{figure}

In order to corroborate the agreement between the number counts over the 2dFGRS strips and models 
guided by the 2dFGRS $n(z)$ \citep{fri}, we have determined the number counts in the $B$-band from the 2dFGRS 
parent catalogue and the $K$-band from 2MASS full release data. These are shown in Figs.~\ref{fig:2dfngp} 
and ~\ref{fig:2dfsgp} for an approximately equivalent magnitude range     
for the $\approx$400 deg$^2$ 2dFGRS Northern declination field and $\approx$600 deg$^2$ 
2dFGRS Southern field respectively, together with models incorporating the corresponding 2dFGRS $n(z)$.
These models are formed through varying the luminosity function parameter $\phi$*, such that the luminosity function 
normalisation varies as a function of redshift to match the observed 2dFGRS $n(z)$. These models are described in 
more detail by \citet{fri}.

The counts are in good agreement with these models and confirm
the earlier work of \citet{fri} suggesting that features in the bright number counts are exclusively due 
to real structures in the local galaxy distribution. The fact that the deficit in the counts can be explained purely by features
in the $n(z)$ and that this agreement is present to the same extent in different passbands suggests that deficiencies in bright
magnitude galaxy counts are unlikely to be caused by strong luminosity evolution at low redshifts.

Having verified the usefulness of number counts as a probe of the local galaxy distribution, we now wish 
to examine the counts over the much larger APM survey area in order to determine to what extent the 
observed deficiency in the optical APM counts may be due to a real underdensity, and whether the deficit in the 
optical counts is complemented by a similar deficit in the 2MASS $K$-band counts.

\begin{figure}[!ht]
\centerline{\epsfxsize = 3.0in
\epsfbox{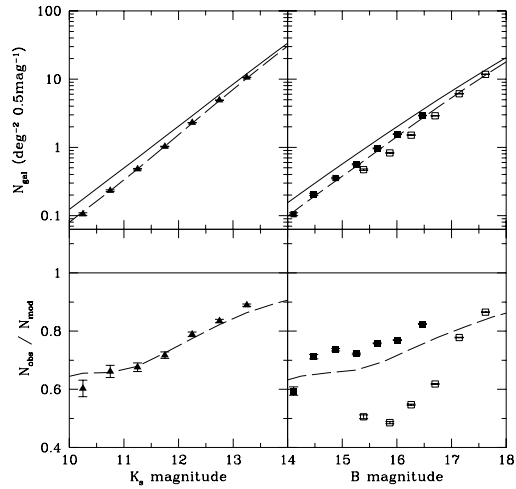}}
\caption{Number counts for the APM field as in Fig.~\ref{fig:2dfsgp}. In the $B$-band plots, the \citet{met}
corrected APM counts are indicated by the open squares, with the new \citet{bus} corrected points by the solid
squares. The dashed line indicates the model incorporating the 2dFGRS Southern $n(z)$.}
\label{fig:apm}
\end{figure}

In Fig.~\ref{fig:apm} we show the optical and $K$-band counts for the $\approx$4000 deg$^2$ APM survey area. In the optical 
count plots we show both the \citet{met} corrected APM survey area counts and also the new counts incorporating the 0.31 
magnitude zeropoint correction to the $B<$17.15 APM photometry determined by \citet{bus}. We also include the models 
incorporating the 2dFGRS Southern $n(z)$ as in Fig.~\ref{fig:2dfsgp}. Surprisingly, the $K$-band and the corrected $B$-band 
counts are in reasonable agreement with these models. This suggests that the form of the galaxy distribution observed in the 
2dFGRS Southern field (including a $\approx$30\% deficiency to $z$=0.1) may extend over the entire APM field of $\approx$4000 
deg$^2$, implying a deficiency of $\approx$30\% over $\approx$1.2$\times$10$^7$ $h^{-3}$Mpc$^3$. This is supported by other 
redshift surveys over the APM field including the LCRS, D-UKST and ESP which indicate similar deficiencies over 
smaller solid angles.

\section{The Significance of the APM Under-Density}

In order to investigate the cosmological relevance of this under-density, 
we now wish to determine the significance of such a feature using assumed 
forms of the two point correlation function. We use the 3-dimensional
analogue of equation 36.6 in \citet{peb}:

\begin{equation}
\left(\frac{\delta N}{\bar{N}}\right)^2=\frac{1}{\bar{N}}+\frac{1}{V^2}\int_0^{r_{cut}}dV_1dV_2\xi (r_{12})
\end{equation}
         
\noindent where $\xi (r_{12})$ is the value of the two-point correlation function between two volume elements
$dV_1$ and $dV_2$. $V$ is the entire volume of the survey, $n$ is the mean galaxy density such that $\bar{N}=nV$ is
the expected number of galaxies in the survey volume. In the first instance, we take a power law form for the correlation function 
($\xi=\left(\frac{r}{r_0}\right)^{\gamma}$) to a limiting distance $r_{cut}$, beyond which the clustering is
assumed to be insignificantly small.

For the APM underdensity, we assume a deficiency in the galaxy distribution of 25\% to $r=300$\mpc\  
($z$=0.1) over $\approx$4000 deg$^2$. We use both the 2dFGRS \citep[][$\gamma$=-1.67]{haw} and $\Lambda$CDM (from the 
Hubble Volume mock catalogue, Carlton Baugh - priv. comm.) real-space correlation functions. The corresponding 
significance estimates are shown in Table~\ref{table:sig}.

\begin{table}
\centering
\begin{tabular}{||c|c|c||} \hline
         $\xi$    & $r_{cut}$ (\mpc\ )    & significance  \\ \hline 
         2dFGRS   & 40                   & 6.0$\sigma$   \\
         2dFGRS   & 300                  & 2.8$\sigma$   \\
         $\Lambda$CDM   & 300                  & 4.6$\sigma$   \\ \hline
\end{tabular}
\caption{\small{Significances for the APM underdensity using the 2dFGRS and $\Lambda$CDM (determined from 
the Hubble Volume mock catalogue) real-space correlation functions. The power law form of the correlation 
function is assumed in each case to a limiting distance $r_{cut}$ beyond which the clustering is assumed to be negligible. }}
\label{table:sig}
\end{table}

Since the real-space correlation function is poorly constrained beyond $r\approx$40\mpc, we take the form of the 2dFGRS 
correlation function beyond this scale to two extremes; firstly, that there is no structure ($r_{cut}$=40\mpc\ ), and secondly 
that the power law form continues effectively to infinity ($r_{cut}$=300\mpc\ ). The Hubble Volume mock catalogue real-space 
correlation function has a break at $r\approx$40\mpc, beyond which the slope is much steeper. Only in the extreme case where the 
power law extends to $r\approx$300\mpc\  does the APM deficiency start to be explained by the correlation function. This indicates 
that the local hole can only be explained if the large-scale correlation of galaxies is significantly more extended than detected 
by the 2dFGRS or predicted by the $\Lambda$CDM model.

We can also gain a handle on the significance of the under-density more directly by using the 
2-dimensional equivalent of equation 1 which employs the 2-point angular correlation function. 
We take the $observed$ $K_s<$12.5 deficiency in the 2MASS $K$-band number counts over the APM survey area of 25\% 
and the best fit angular correlation function for $K_s<$12.5 Hubble Volume mocks \citep[a two power law fit with a break at 
10$^{\circ}$,][]{fri1}. This magnitude limit is chosen to match the peak in the selection function to the redshift range of 
interest. The implied significance with this two power law form for the angular correlation function (and no cut) is 5.0$\sigma$.

\section{The 2MASS Galaxy Angular Power Spectrum}

The under-density in the APM survey area described in section 2 may imply the presence of excess power at large scales over that 
seen in the 2dFGRS or the $\Lambda$CDM Hubble Volume. In order to try and probe the form of the 
clustering at extremely large scales we compute the angular power spectrum of 2MASS galaxies. The angular power spectrum is useful 
in determining the form of galaxy fluctuations at extremely large scales. The 2MASS survey is particularly useful for applying 
this technique as it covers almost the entire sky and so has a well-behaved window function.

We compute the angular power spectrum for $K_s<$13.5 (corrected for extinction), $b\ge$20$^{\circ}$ 2MASS galaxies using a 
spherical harmonic expansion of the number density:

\begin{equation} 
\sigma (\theta ,\phi ) =  \sum_{l} \sum_{m} a_l^m Y_l^m(\theta ,\phi )
\end{equation}

\noindent \citep{peb2,sch} The coefficients of this expansion, $a_l^m$, are estimated over the observed solid angle
$\Omega_{obs}$:

\begin{equation}
a_l^m =\sum_{N_{gal}} Y_l^m(\theta ,\phi ) - {\mathcal{N}} \int_{\Omega_{obs}} Y_l^m(\theta
,\phi ) d\Omega
\end{equation}

\noindent where ${\mathcal{N}}$=$N_{gal}/\Omega_{obs}$ is the observed number of galaxies per steradian. The angular power is then
determined as the ratio of the observed signal to the expected Poisson fluctuation, ${\mathcal{N}}$:   

\begin{equation}
C_l = \frac{1}{{\mathcal{N}}(\Omega_{obs}/4\pi)(2l+1)} \sum_m |a_l^m|^2
\end{equation}

\noindent such that $C_l$=1 corresponds to a random distribution for all-sky coverage.

In order to compare the empirical power spectrum to some cosmological prediction, we take the transfer function fitting 
formulae of \citet{eis} which are essentially perfect in the linear regime ($l$\lsim30). We take a linear biasing scheme such 
that $P_{gal}(k)$=$b^2P_{matter}(k)$. The three dimensional power spectrum is then collapsed to two dimensions using a Bessel 
function transform, and the 2MASS selection function is imposed upon it \citep{sch}.

\begin{figure}[!ht]
\centerline{\epsfxsize = 3.4in
\epsfbox{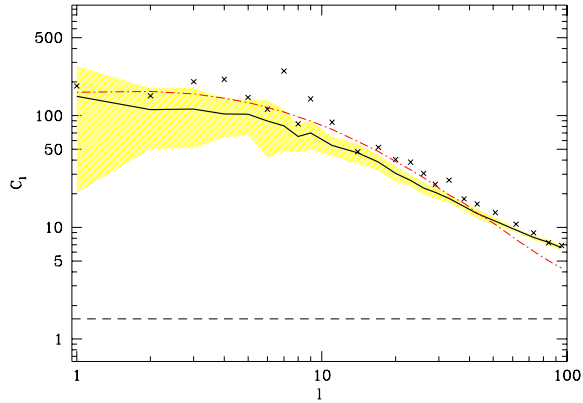}}
\caption{The $K_s<$13.5, $|b|>$20$^{\circ}$ 2MASS galaxy angular power spectrum (crosses) are shown with the mean and
1$\sigma$ spread from the power spectra from 27 mock 2MASS catalogues constructed from the $\Lambda$CDM Hubble Volume
mock catalogue. The dot-dashed line indicates the predicted linear form to the mock 2MASS power spectra using the \citet{eis}
transfer fitting formulae for the Hubble Volume mock catalogue input parameters of $\Omega_m$=0.3, $\Omega_b$=0.04, $h$=0.7 and
$\sigma_8$=0.9. The dashed line indicates the expected Poisson fluctuation for the solid angle used.}
\label{fig:cl1} 
\end{figure}

\begin{figure}[!ht]
\centerline{\epsfxsize = 3.4in
\epsfbox{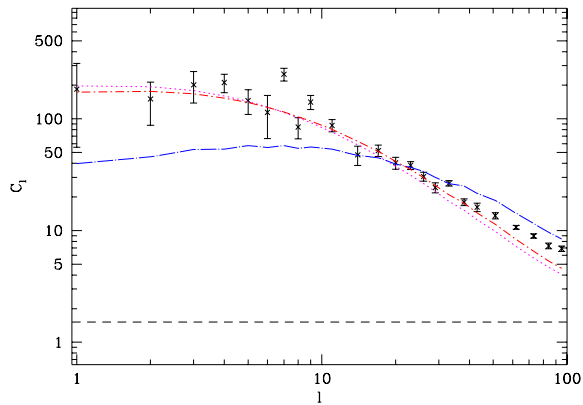}}
\caption{Here we show the 2MASS galaxy angular power spectrum as in the previous figure, with errorbars taken from the 1$\sigma$
spread of the 27 mock 2MASS catalogue power spectra. This is compared to the predicted linear form of the angular power spectrum
for the Hubble Volume mock input parameters as before (dot-dashed line), and also to the best-fit WMAP-2dFGRS
parameters (solid line) of $\Omega_m$=0.27, $\Omega_b$=0.044, $h$=0.71 and $\sigma_8$=0.84 \citep{ben} and an SCDM prediction
using $\Omega_m$=1.0, $\Omega_b$=0.04, $h$=0.5 and $\sigma_8$=0.84 (large dot-dashed line). In each case we apply a bias of 
$b$=1.1.}
\label{fig:cl2}
\end{figure}

In Fig.~\ref{fig:cl1} we show the angular power spectrum for 518,576 $K_s<13.5$, $|b|>$20$^{\circ}$ 2MASS galaxies. Also shown are 
the mean and 1$\sigma$ spread from 27 mock 2MASS catalogues constructed from the $\Lambda$CDM Hubble Volume mock catalogue, and 
the corresponding linear model for the input Hubble Volume mock catalogue parameters. In each case, we take a bias of $b$=1.0. We 
compute the angular power spectrum to $l$=100 only, since beyond this the angular scales involved become dominated by uncertain 
forms to the bias.

The 2MASS angular power spectrum is in reasonable agreement with the unbiased mock angular power spectra although the 2MASS 
angular power spectrum slope is steeper, implying some discrepancy with the Hubble Volume mock in the unbiased 
case. Therefore it seems that either the Hubble Volume mock angular power spectrum does not agree with the 
2MASS angular power spectrum or there exists a scale-dependent bias. With the introduction of scale-independent bias this 
discrepancy may clearly shift to smaller scales where there is good agreement in the unbiased case; clearly, the issue of bias 
is critical in determining the level of disagreement between the Hubble Volume mock catalogue and 2MASS. If we take a bias of 
$b=1.1$ \citep{mal}, the disagreement at large scales ($l\le$30) is at the $\approx$3$\sigma$ level. The model linear 
prediction is in good agreement with the mock 2MASS power spectra below $l\approx$30 although the model slope is 
shallower than the mean mock power spectrum; beyond this scale non-linear effects begin to dominate.

We wish to probe to what extent the 2MASS angular power spectrum discriminates between different cosmologies, and whether it is 
in disagreement with $\Lambda$CDM. In Fig.~\ref{fig:cl1} we show the 2MASS angular power spectrum as before with errorbars taken 
from the 1$\sigma$ spread of the 27 mock 2MASS power spectra; in doing this we assume that the form of the clustering in the 
Hubble Volume mocks is realistic at the scales of interest. Along with this we show a $\Lambda$CDM linear prediction using the 
best-fit WMAP-2dFGRS parameters, the Hubble Volume input parameter prediction as before, and an SCDM prediction. To these models 
we apply a linear bias of $b$=1.1 \citep{mal,fri1}. 

Using this bias prescription, the $\Lambda$CDM linear predictions derived from the WMAP-2dFGRS best-fit and the Hubble Volume mock 
input parameters are in excellent agreement with the 2MASS angular power spectrum to $l\approx$40. The SCDM prediction gives a 
very poor fit to the 2MASS angular power spectrum, indicating that this analysis may provide useful constraints on the cosmology; 
this is examined by \citet{fri1}.  

While there appears to be an excess of power in the 2MASS galaxy angular power spectrum compared with the $\Lambda$CDM Hubble 
Volume mocks at large scales, it is not enough to account for the local hole described in sections 2 and 3: We use a Bessel 
function transform to convert the angular power spectrum to the 2-point angular correlation function; the result is in good 
agreement with the 2MASS correlation function determined by \citet{mal}, and so the implied significance of the local hole 
described in sections 2 and 3 remains high. Therefore, somewhat paradoxically, the local hole in the 2MASS data in the APM survey 
area does not seem to be explained by the angular power spectrum or 2-point correlation functions from the same 2MASS data; 
possible explanations are discussed below.

\section{Summary}

In order to investigate the number count deficiency in the SGC detected by the APM survey, we have used models guided by the 
2dFGRS $n(z)$ as predictions of the $B$-band and $K$-band number counts. First, we wanted to verify the usefulness of this 
technique as a probe of the underlying large-scale structure and so compared the 2dFGRS $n(z)$ models to the corresponding 
number counts extracted from the 2dFGRS parent catalogue and the 2MASS full release catalogue for the 2dFGRS fields. In both 
passbands and both the Northern and Southern declination fields, the predicted counts agree very well with the observed counts. 

Using this technique, we then probed the galaxy distribution over the much larger APM field of $\approx$4000 deg$^2$. The 2MASS 
counts and the APM counts (corrected for a zero-point offset of 0.31 magnitudes found with respect to recent $B$-band CCD 
photometry \citep{bus}) are in good agreement, and suprisingly, agree well with the 2dFGRS Southern $n(z)$ model. This suggests 
that a similar deficiency to the 2dFGRS Southern distribution exists over the entire APM field; that would be a $\approx$30\% 
deficiency to $z\approx$0.1 over $\approx$4000 deg$^2$.

This deficiency, while less extreme than the galaxy distribution implied by the original APM counts, poses a 
significant problem for our present understanding of clustering and possibly a $\Lambda$CDM cosmology. We calculate 
the significance of such an underdensity, considering several possible forms for the 
real-space correlation function. In all cases the APM deficiency as described above represents a \gsim3$\sigma$ fluctuation. In 
particular we found that the APM deficiency represents a 4.6$\sigma$ fluctuation using the $\Lambda$CDM Hubble Volume mock 
correlation function.

The deficiency in the APM survey area therefore suggests an excess of power which
is incompatible with the 2dFGRS or Hubble Volume mock $b$=1 $\Lambda$CDM  
angular correlation functions; only an extension of the small scale 2dFGRS
$\gamma$=-1.67 power law correlation function would explain the hole. In
order to investigate this discrepancy and probe the galaxy clustering at
extremely large scales, we finally determined the angular power spectrum 
of 2MASS $K_s<13.5$, $|b|>20^\circ$ galaxies and also 27 mock 2MASS
catalogues constructed from the $\Lambda$CDM Hubble Volume mock catalogue.
There is a discrepancy between the two in the range $1<l<100$, in that the
Hubble Volume power spectrum slope is significantly flatter. Inclusion of linear bias
will therefore not improve this agreement. However, we find that the     
increased power at large scales is still not large enough to explain the
deficiency in the APM survey area.  In the more restricted range of $1<l<30$, 
where the effect of non-linear clustering and any scale dependent bias may
be assumed to be less, $\Lambda$CDM models can give a good fit to the
2MASS power spectrum and the constraints on the cosmological parameters
from fitting these data are explored by \citet{fri1}.

In conclusion, the issue of the large local hole has yet to be resolved.
The new data from the 2dFGRS and 2MASS catalogues along with the large
$B$-band CCD survey of \citet{bus} indicates that the low APM
counts may be due partly to a zero-point error in the bright APM
photometry and partly to a large deficiency in the galaxy distribution
stretching to $z$=0.1. The angular power spectrum of the 2MASS galaxies, although
steeper than the Hubble Volume $\Lambda$CDM mock catalogue, still has too
shallow a slope to explain the hole. This may not be too surprising since
the depth of 2MASS is comparable to the previous $B<$17 surveys in the APM survey 
area which only sampled the deficient volume with $z<0.1$. So it is still
possible that even the entire 2MASS volume may still not represent a fair
sample of the Universe. Another possible explanation is that the
clustering is non-Gaussian on the scales of 150-300$h^{-1}$Mpc. Although
this seems unlikely, there have recently been tentative reports of non-Gaussianity
detected on even larger scales in the WMAP CMB data \citep[e.g.][]{eri}. 
The final possibility is that the agreement between the $B$ and $K$-band 
counts predicted on the basis of 2dFGRS $n(z)$ models and the data is
coincidental and this has caused us to overestimate the angular size of
the deficiency in the APM survey area. A new $B<19.5$ galaxy redshift survey
over the whole 4000deg$^2$ APM survey area would help resolve this question.


\begin{thebibliography}{}
\bibitem[\protect\citeauthoryear{Bennett et al.}{2003}]{ben} Bennett, C.L. et al. 2003,  ApJS, 148, 1
\bibitem[\protect\citeauthoryear{Busswell et al.}{2004}]{bus} Busswell, G.S., Shanks, T., Outram, P.J., Frith, W.J.,
Metcalfe, N. \& Fong, R. 2004, accepted by MNRAS, astro-ph/0302330
\bibitem[\protect\citeauthoryear{Colless et al.}{2003}]{col} Colless, S. et al. 2003, astro-ph/0306581
\bibitem[\protect\citeauthoryear{Eisenstein \& Hu}{1998}]{eis} Eisenstein, D.J. \& Hu, W. 
1998, ApJ, 496, 605
\bibitem[\protect\citeauthoryear{Eriksen et al.}{2004}]{eri} Eriksen, H.K., Novikov, D.I., Lilje, P.B., Banday, A.J. \& Gorski, 
K.M. 2004, astro-ph/0401276
\bibitem[\protect\citeauthoryear{Frith et al.}{2003}]{fri} Frith, W.J., Busswell, G.S., Fong, R., Metcalfe, N.
\& Shanks, T. 2003, MNRAS, 345, 1049
\bibitem[\protect\citeauthoryear{Frith et al.}{2004}]{fri1} Frith, W.J., Shanks, T. \& Outram, P.J. 2004, in preparation   
\bibitem[\protect\citeauthoryear{Hawkins et al.}{2003}]{haw} Hawkins, E. et al. 2003,  MNRAS, 346, 78
\bibitem[\protect\citeauthoryear{Jarrett et al.}{2000}]{jar} Jarrett, T.H., Chester, T., Cutri, R., Schneider, S.,
Skrutskie, M. \& Huchra, J.P. 2000, AJ, 119, 2498
\bibitem[\protect\citeauthoryear{Maller et al.}{2003}]{mal} Maller, A.H., McIntosh, D.H., Katz, N. \& Weinberg, M.
2003, astro-ph/0304005
\bibitem[\protect\citeauthoryear{Maddox et al.}{1990}]{mad} Maddox, S.J., Sutherland, 
W.J., Efstathiou, G. \& Loveday, J. 1990, MNRAS, 243, 692
\bibitem[\protect\citeauthoryear{Metcalfe et al.}{1995}]{met} Metcalfe, N., Fong, R. \&  
Shanks, T. 1995, MNRAS, 274, 769
\bibitem[\protect\citeauthoryear{Peebles}{1973}]{peb2} Peebles, P.J.E.. 1973, ApJ, 185, 413
\bibitem[\protect\citeauthoryear{Peebles}{1980}]{peb} Peebles, P.J.E. 1993, Principles of Physical Cosmology,
Princeton University Press
\bibitem[\protect\citeauthoryear{Ratcliffe et al.}{1998}]{rat} Ratcliffe, A. et al. 1998, MNRAS, 300, 417
\bibitem[\protect\citeauthoryear{Scharf et al.}{1992}]{sch} Scharf, C., Hoffman, Y., Lahav, O. \&  Lynden-Bell, D.
1992, MNRAS, 256, 229
\bibitem[\protect\citeauthoryear{Shectman et al.}{1996}]{she} Shectman, S.A., Landy, S.D., Oemler, A., Tucker, D.L., Lin, H., 
Kirshner, R.P. \&  Schechter, P.L. 1996, ApJ, 470, 172
\bibitem[\protect\citeauthoryear{Vettolani et al.}{1997}]{vet} Vettolani, G. et al. 1997, A\&A, 325, 954
\end{thebibliography}
\end{document}